# Capturing the Connections: Unboxing Internet of Things Devices

**Kami Vaniea**
School of Informatics
Informatics Forum 5.23
10 Crichton Street
University of Edinburgh

**Ella Tallyn**
Design Informatics
78 West Port, Edinburgh
University of Edinburgh

**Chris Speed**
Design Informatics
78 West Port, Edinburgh
University of Edinburgh

**Abstract**
Based upon a study of how to capture data from Internet of Things (IoT) devices, this paper explores the challenges for data centric design ethnography. Often purchased to perform specific tasks, IoT devices exist in a complex ecosystem. This paper describes a study that used a variety of methods to capture the interactions an IoT device engaged in when it was first setup. The complexity of the study that is explored through the annotated documentation across video and router activity, presents the ethnographic challenges that designers face in an age of connected things.

**Introduction**
This pictorial presents an ethnographic method used to study the behaviours of an Internet of Things device during its initial setup and use. The work represents an ethnographic approach to a common computer security practice of observing IoT devices as they are unboxed and begin interacting with the world around them. The challenge here is of capturing both the invisible and rapid, data communications between devices, as well as the observable communications with the user via screen interfaces.

The field of IoT is at a stage of rapid development, and many new devices designed for the domestic environment, for example: smart meters, kettles, bathroom scales and plug sockets, are taking up residence in our homes. This rapid development creates challenges in integrating these devices into home settings where everything from the user interface to the security implications can impact the devices' acceptability. Within IoT there has been considerable innovation in the development of novel user experiences, in which methods to operate and otherwise interact with devices have been explored, for example gestural input as in Moen's hand free faucet [11] or remote controls using a smartphone app as in the WeMo plug studied here.

The combined problem for both ethnographers and security specialists studying IoT is creating an accurate and holistic recording of how the device interacts with everything around it. The work presented here represents a collaboration between authors with ethnographic and security backgrounds working together to address the problem through the development of an ethnographic method which builds on the experience of both groups.

## Security

Security and ultimately user acceptance are also concerns as vendors tend to either have no security at all or poorly designed security [1]. The news regularly informs consumers how their kettles leak passcodes [8] and dolls record children's voices [9]. Consequently users are concerned about inviting these devices into their homes and need more effective methods of understanding their behaviors [5].

When examining a new device, security researchers need to capture the technical interactions in great detail. A device may perform an important activity, such as scanning a network, only once. The device may also change its own code on first use by updating the software. Once these events occur the original state of the device may be unrecoverable even with a reset.

## Ethnography

Ethnographic studies in HCI aim to reveal the rich details of experiences and interactions between complex computational interfaces, people and settings. This is done through recording and analysis of interactions taking place, so that they can be clearly seen and understood, in particular by clarifying the relationships and exchanges between various actors and their environment [6]. However, most ethnographic studies do not record the details of the invisible system communications that are occurring at the same time as the human interactions. An exception was the recording of human physical activity in outdoor locations that is synchronised with recordings of computational actions and interactions has been attempted in multi-participant, distributed, outdoor experiences [7]. This synchronised data has provided a rich picture of events that spans the physical and digital, which can be interrogated and interpreted by ethnographers and designers, and reveals the complex and sometimes surprising relationships and dependences between the digital and physical.

## The Brief

Hacking or breaking into the device can cause changes to its behavior making IoT devices similar to a human ethnographic subject, whose mind cannot be read, and, as in strict naturalistic studies we cannot interrogate without affecting natural behaviours [3]. In this work we therefore treat the IoT device as we would a human subject [12], recording it from multiple angles in an attempt to understand how it naturally interacts with its environment.

To illustrate the approach we present 110 seconds of interaction with a WeMo Insight Switch developed by Belkin (a plug socket with a digital switch).

The recording setup pictured in Figure 1 contains the following recording equipment: 1. two autographers one around the neck of the researcher and the other near or on the IoT device. 2. A web camera pointed down at the mobile phone. 3. A camcorder on a tripod setup to capture the whole table. 4. The router which is setup to capture all traffic coming across it (located near network plug and not on table). 5. Screencapture from the mobile device of everything shown on the screen.

Non-recording equipment included: A. a standard table lamp to support the webcam over the phone. B. A power plug for the IoT device that is visible from the main camera. C. A light bar, plugged into the WeMo, to make it highly apparent when the WeMo is off or on and also to use electricity in case that impacts its behavior. D. A set of index cards (not pictured) which were shown to the camera when the researcher switched tasks. E. The box the IoT device came in (not pictured) which is visible to the camera during unboxing. F. Any instructions that came with the device (not pictured), also used during the setup. G. A checklist of all the different recordings the researcher needed to initiate in the order they should be initiated (not pictured). For example, the main camera needs to start first then the other recording so that the main video can be used to sync the other recordings.

## The setup

This setup was designed to capture the unboxing and initial use of an IoT device in a controlled environment so as to record the event from as many perspectives as possible. Since most IoT devices lack keyboards and screens the setup process often involves either an app on a mobile device or a web page. The actions of the user are also important to observe since the human may be interacting across devices or reacting to information not obviously apparent such as an included instructions sheet. Hence the recordings focus on three subjects: the IoT device, the accompanying app (if any) and the human.

Laptop to monitor position of phone under camera, connections on router and status of router recording

Video still from main camera

Image from Autographer on Wemo

TCP connection stream between Phone and WeMo as seen by the router

Image from Autographer on human

Video still from webcam

Screen capture still from phone

# A moment in time

The main camera, webcam and two autographers capture the human-visible parts of the scene in high detail and show i) where the human is looking, and ii) where their hand is relative to the device. The phone collects a screencapture of the phone interface which avoids occlusion of the interface by the researcher's hand. The router is collecting communications using the standard tcpdump [10] program. Finally, the laptop is used by the researcher to initiate recordings, make sure the phone is in view of the webcam, and monitor the router.

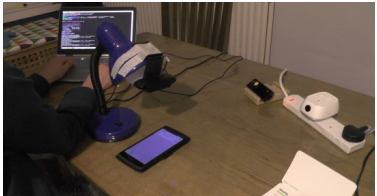
Turn on packet capture

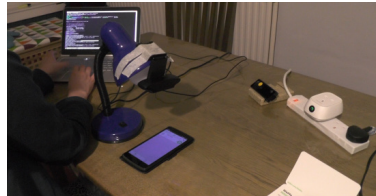
Check that packet capture is working

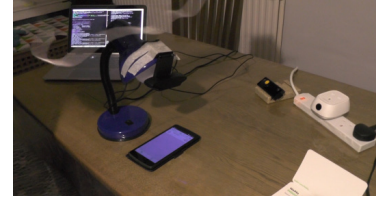
Realise that small light on Wemo is not visible to the main camera, and go and fetch bigger light strip

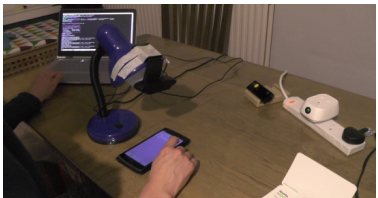
Push 'on' on Phone app

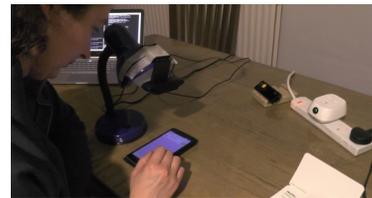
Push 'on' on phone app

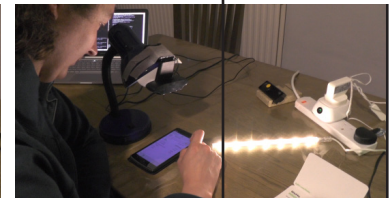
Push 'on' on phone app

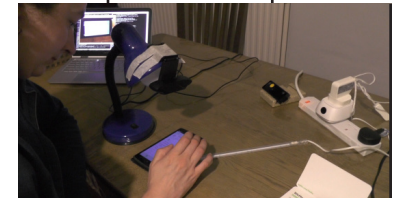
Push 'off' on phone app

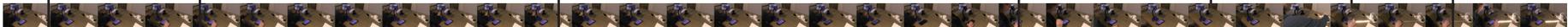
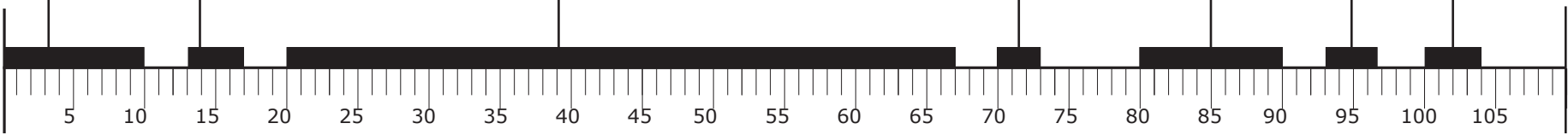

## Point of View: Human

From a user's point of view (main camera) the WeMo is a small box plugged into a power strip which can be controlled by an app. The switch is seen to be on or off based on the light on the front or the output of the app. The indications of the WeMo status are so subtle that the researcher fetches a light bar and plugs it into the WeMo to make the WeMo's state more apparent. The diagram shows the interaction between the researcher and WeMo from the researcher's perspective.

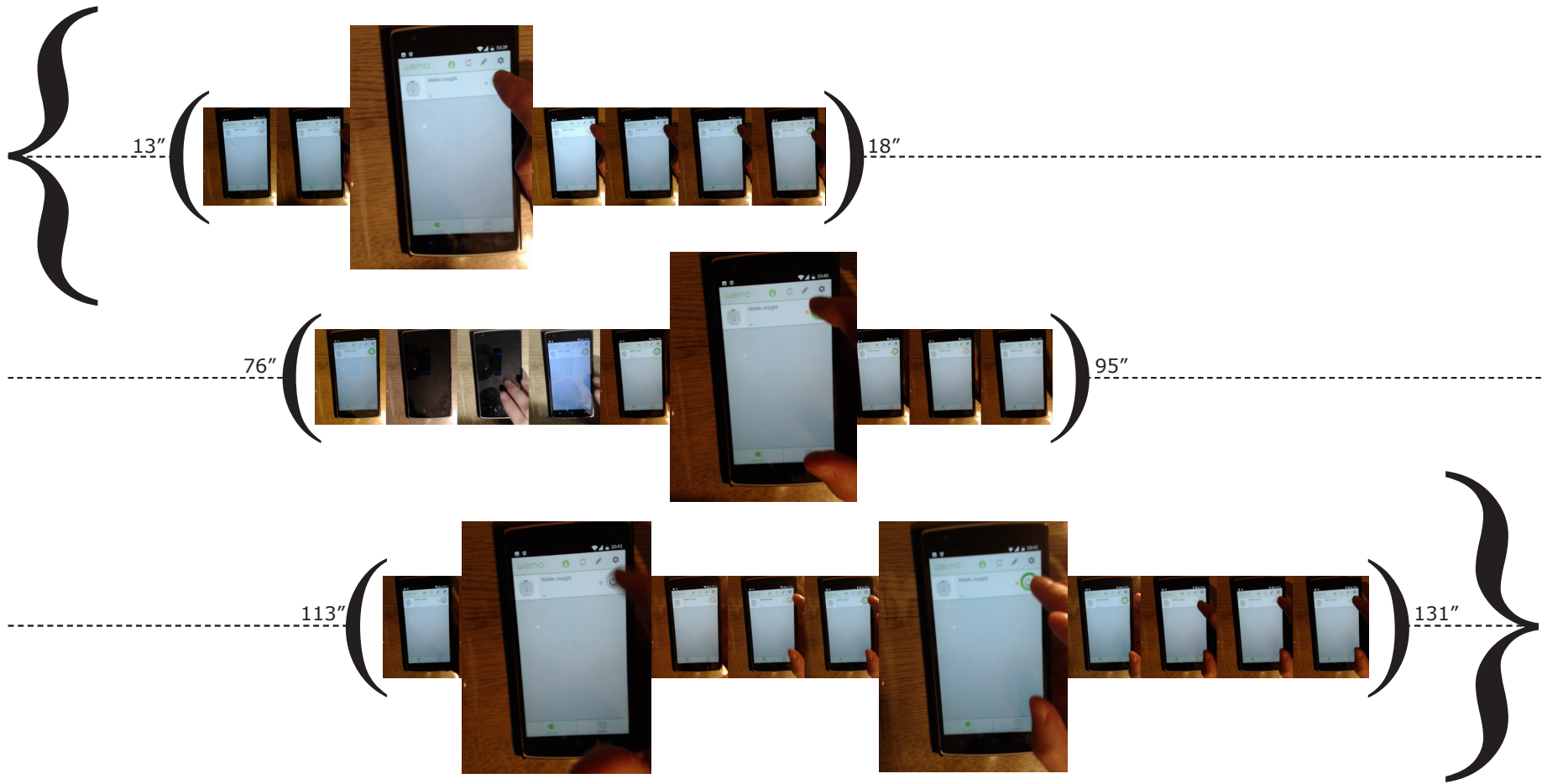

## Point of View: Phone

The WeMo is intended to be interacted with through the provided phone app making the app a central point of the interaction. In its default view the app has very little detail primarily providing a power button which shows the states of: off (grey), on (green), and transition (orange).

Consequently the webcam's point of view is rather simplistic and limited to only the state of WeMo and the position of the researcher's hand.

Pictorials

## Router conversation log

**Router** Yes — You there? **Wemo**
**Phone** Turn on please
 Got it **Wemo**
 I'm on now **Wemo**
 Let's start an encrypted chat **Wemo**
**api.xbcs.net** Sure
 akdfjw;eifjwe;ij **Wemo**
**api.xbcs.net** dslkfs;ldkfjs
 Anyone have a plug and play service they are sharing? **Wemo**
 I hear you are sharing some services **Wemo**
**Windows** Yep, here is a full list ....
 You there? **Wemo**
**Router** Yes
**Phone** can you send me the setup data file?
 Sure, here you go **Wemo**

**Phone** can you send me your current status?
 Sure, here you go **Wemo**
**Phone** Can you send me your power settings for this home?
 Sure, here you go **Wemo**
**Phone** Can you send data you have collected so far?
 Sure, here you go **Wemo**
**Phone** What time do you think it is right now?
 Sun, 29 May 2016 224035 GMT **Wemo**
**Phone** If any events happen to you in the future please tell me about them.
 OK **Wemo**
**Phone** If you want to update please tell me that too.
 OK **Wemo**
**Phone** Just making sure you know I am allowed to turn you on and off.

 Understood, here are all the secret codes **Wemo**
 Let's start an encrypted chat **Wemo**
**api.xbcs.net** Sure
 fewkfbjek>?33 **Wemo**
**api.xbcs.net** oerle,,ewjd
**Phone** Anyone support Belkin protocols?
 Remember how you told me to tell you about any events? Some data about my plugins just changed. **Wemo**
**Phone** OK
 Oh, my home and device ID codes just changed too. **Wemo**
**Phone** OK
 I've got a problem, I'm behind a home network router so if the phone I have been talking to leaves home I won't be able to talk to it anymore. Could you act as a mediary and connect us if that happens? **Wemo**
**Belkin** Sure

 Let's start an encrypted chat **Wemo**
**api.xbcs.net** Sure
 duorllwgar>?d **Wemo**
**api.xbcs.net** []d,mwhh,e
 []d,mwhh,e **Wemo**
**Phone** Turn off please
 Got it **Wemo**
 I'm off now **Wemo**
 Let's start an encrypted chat **Wemo**
**api.xbcs.net** Sure
 iotowy;;s.gppr **Wemo**
**api.xbcs.net** smnvoeuu[
 Hello bug reporting server, lets start an ecrypted chat. **Wemo**
**Bugsense** Sure
 iotowy;;s.gppr **Wemo**
**Bugsense** smnvoeuu[
**Phone** Turn on please
 Got it **Wemo**
 I'm on now **Wemo**
**Phone** Turn off please
 Got it **Wemo**
 I'm off now **Wemo**

## Point of View: Router

The router sees the detailed communications between devices. Most of those communications are not encrypted meaning that the router can listen into these conversations. Not encrypting these types of communications is a common issue in IoT devices. We translate the technical communication data into a chat-style communication using the content and protocol information to convey the nature of the conversation. During the 110 seconds, the WeMo has 31 conversations with 7 partners, 3 of which are outside the home.

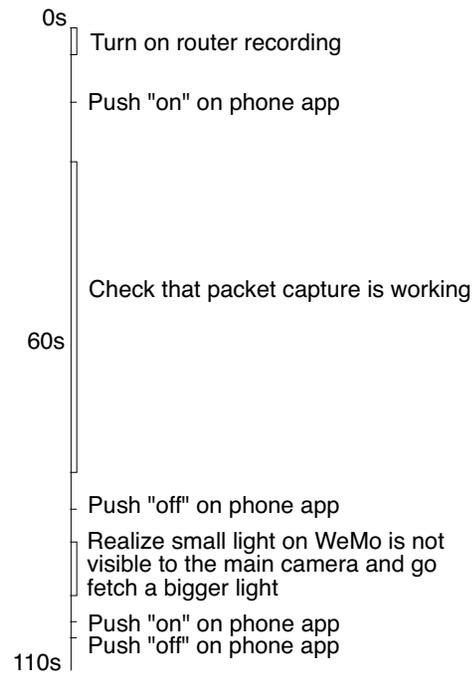
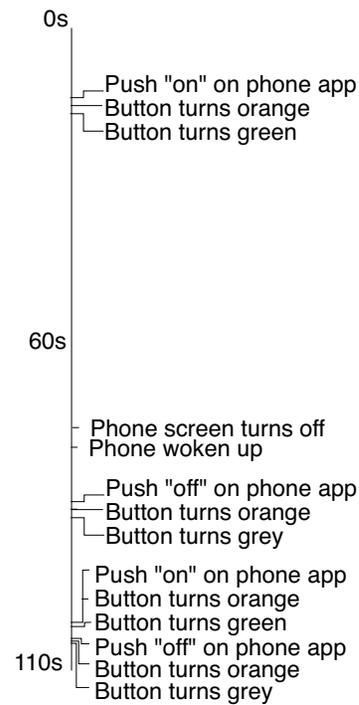
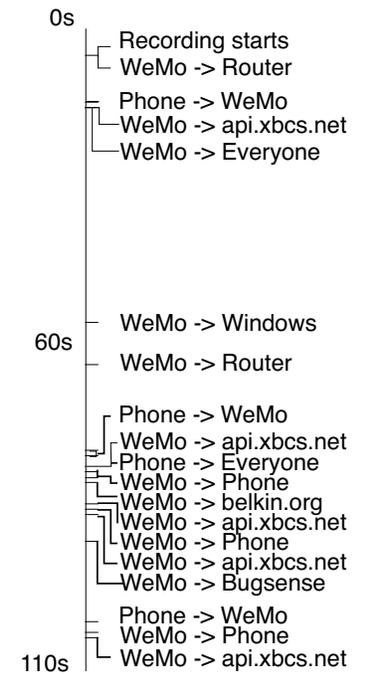

## Point of View: Holistic

Putting the timelines together, the most striking observation is that the only event visible across all three timelines is when the user clicks the power button on the phone app. Doing so is visible from the main camera, webcam, and generates a message that the router can see. All other activities are invisible from at least one of the points of view. For example, the router and webcam cannot see that the user has left the table. Similarly the main camera and webcam cannot see when the WeMo contacts the belkin.org web server.

### Connecting the views
Logically a user might expect that the WeMo can only be in one of two states: on or off. However, the phone UI has a third state: orange. The main camera captures the user's confusion over the orange button as they glance up at the WeMo itself, the webcam also sees the user's finger hovering uncertainly over the button. The router communication however explains the situation detailing how the phone asks the WeMo to turn on which is immediately acknowledged but it isn't till a full second later that the WeMo reports successfully turning on, which causes the phone to change the button's color to green.

The interactions between human, phone, and router are also surprisingly human driven. One might expect the WeMo to be chattering away at all times to other network devices. However, the prior figure instead shows that its communications are clustered around human events such as pushing the power button or waking up the phone. When the phone is woken up it immediately starts interrogating the WeMo for all sorts of status information in case the user might want it. After each on/off event the WeMo contacts api.xbcs.net using an encrypted communication. When the user plugs the light bar into the WeMo an encrypted communication is started with Bugsense an online code bug reporting service. This suggests that plugging in the lightbar may have caused an error in the WeMo which was invisible to the user.

### Discussion
Ease of use is still the primary goal of interaction design in IoT devices and, in order to make the devices appear simplistic, much of the complex behaviours of the devices is necessarily hidden. However, their behaviours are not simplistic, and this could become increasingly problematic if lack of user unawareness leads to disengagement with potential security implications. Blythe and Lefevre have suggested that users of new IoT devices need to take more care of cyberhygiene, that they may be unaware of risks, and that an attempt should be made to increase user capabilities [4]. It is possible that the lack of engagement and awareness users currently exhibit is partially due to the invisible nature of the actions of IoT devices. It therefore stands that we may need to consider visualising some of these complex behaviours as part of IoT design.

While designing the protocol we learned some useful lessons about the setup. The main camera turned out to be vital, it captures a view of the whole scene and allowed us to sync recordings fairly easily by observing the researcher initiating them on the laptop. We learned early to always turn it on first and off last. The table layout depicted on page 3 was iteratively developed to ensure that all components were visible to the main camera at all times as well as limit occlusion by the researcher. Cards were also added at key points in the protocol to make searching through the video easier. Having a dedicated router running a standard variant of Linux let us use off-the-shelf technologies and not have to invent our own, it also meant that we could isolate the study traffic and not confuse it with external wifi traffic.

### Future Work:
• Create a system that can take the raw data sources required for security analysis and produce outputs that would be readable by designers, for example the router output shown from the point of view of the router could theoretically be generated automatically.
• Further develop the method so that it can be consistently replicated by other researchers and designers

**Acknowledgements**
We thank all the masters students who contributed to the project, in particular: Dian Yordanov, Nikolaos Tsirigotakis, and Martin Kramer. This research is funded in part by the UK National Cyber Security Centre and the Engineering and Physical Sciences Research Council project: PETRAS (EPN02334X1).